\documentclass[aps,prl,reprint]{revtex4-1}
\usepackage{amssymb,amsmath}
\usepackage{graphicx}
\usepackage[english]{babel}
\usepackage{siunitx}
\usepackage{isotope}
\usepackage{braket}
\usepackage{color}
\usepackage{hyperref}

\usepackage[normalem]{ulem}

\begin{document}
\title{
Supersonic Rotation of a Superfluid: A Long-Lived Dynamical Ring
}

\author{Yanliang Guo$^{1,2}$, Romain Dubessy$^{1,2}$, Mathieu de Go{\"e}r de Herve$^{1,2}$, Avinash Kumar$^{2,1}$, Thomas Badr$^{2,1}$, Aur{\'e}lien Perrin$^{2,1}$, Laurent Longchambon$^{1,2}$, and H{\'e}l{\`e}ne Perrin$^{2,1}$}

\affiliation{$^1$Laboratoire de physique des lasers, Universit\'e Paris 13 Sorbonne Paris Cit\'e, 99 avenue J.-B. Cl\'ement, F-93430 Villetaneuse, France}
\affiliation{$^2$LPL, CNRS UMR 7538, 99 avenue J.-B. Cl\'ement, F-93430 Villetaneuse, France}
\date{\today}

\begin{abstract}
We present the experimental realization of a long-lived superfluid flow of a quantum gas rotating in an anharmonic potential, sustained by its own angular momentum. The gas is set into motion by rotating an elliptical deformation of the trap. An evaporation selective in angular momentum yields an acceleration of rotation until the density vanishes at the trap center, resulting in a dynamical ring with $\simeq 350\hbar$ angular momentum per particle. The density profile of the ring corresponds to the one of a quasi two-dimensional superfluid, with a linear velocity reaching Mach 18 and a rotation lasting more than a minute.
\end{abstract}

\maketitle
Superfluidity is a rich quantum dynamical phenomenon \cite{Leggett1999} with striking manifestations such as the existence of a critical velocity for the creation of excitations~\cite{Allum1977} and the appearance of quantized vortices when set into rotation, as observed in liquid helium \cite{Vinen1961} and in dilute Bose-Einstein condensates (BEC)~\cite{Madison2000,AboShaeer2001}. The particular case of a quantum gas rotating at an angular frequency $\Omega$ has especially attracted a lot of theoretical and experimental interest. Indeed, it presents a strong analogy with a quantum system of charged particles in a uniform magnetic field, relevant for condensed matter problems such as type II superconductors or the quantum Hall effect~\cite{Cooper2008,Bloch2012}. 

In a superfluid quantum gas confined in a harmonic trap of radial frequency $\omega_r$, for rotation rates $\Omega\lesssim\omega_r$ a dense triangular array of singly charged vortices establishes. In the limit $\Omega\simeq\omega_r$ the ground state of the system reaches the atomic analog of the lowest Landau level (LLL) relevant in the quantum Hall regime~\cite{Ho2001,Aftalion2005,Bloch2008RMP}. However, reaching the situation $\Omega\geq\omega_r$ is impossible in a purely harmonic trap because the radial effective trapping in the rotating frame vanishes due to the centrifugal potential, leading to the loss of the atoms.

This high rotation regime requires an anharmonic trap to counteract the centrifugal effect. A crucial point of this new situation is that a zero-density area ---a hole--- grows at the trap center above a critical rotation frequency $\Omega_h$~\cite{Fischer2003,Fetter2005}, leading to an annular density profile. 
Above a second threshold $\Omega>\Omega_{gv}$, the gas enters the so-called ``giant vortex'' regime which has attracted a lot of theoretical attention~\cite{Kasamatsu2002,Lundh2002,Fischer2003,Fetter2001,Fetter2005,Rougerie2011}: the vortex cores all migrate close to the depleted central region~\cite{Kasamatsu2002} and for even higher rotation rates the ground state of the system becomes highly correlated~\cite{Cooper2001}. In the intermediate case where $\Omega_h<\Omega<\Omega_{gv}$, the annular gas is expected to display a vortex array in its bulk and exhibit a rich excitation spectrum~\cite{Cozzini2005} which has not been experimentally studied up to now. Moreover, the velocity of the atomic flow is expected to be supersonic~\cite{Kasamatsu2002}, i.e., exceeding by far the speed of sound. Pioneering experiments have tried to generate a ring-shaped flow in a three-dimensional condensate, either approaching $\Omega_h$ from below in an anharmonic trap \cite{Bretin2004} such that no hole could form, or drilling a hole in a rotating gas confined in a harmonic trap by removing atoms with a laser pulse, the system
being strongly out of equilibrium \cite{Engels2003,Simula2005}.

\begin{figure}[t]
\includegraphics[height=3cm]{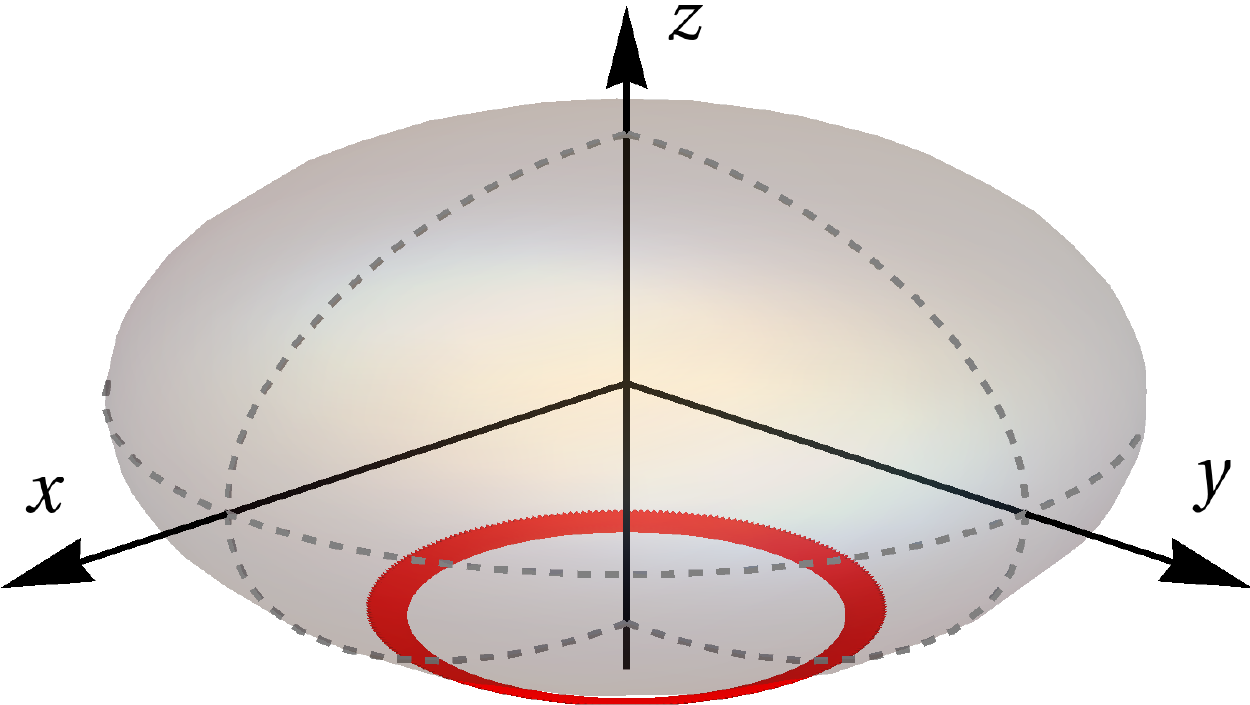}
\includegraphics[height=3cm]{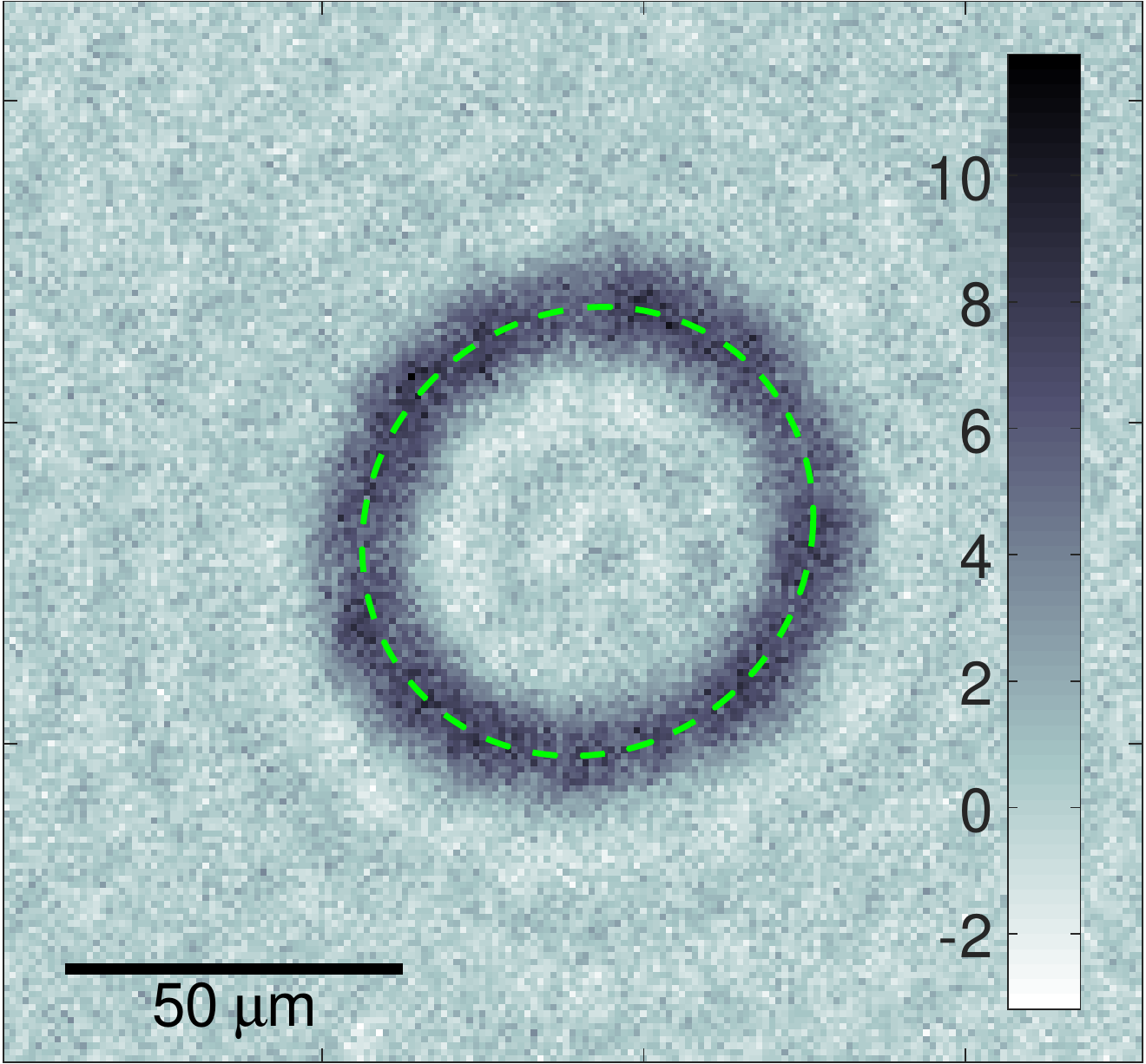}
\caption {\label{ring358}(a) Computed density contour (red annulus) for a BEC rotating at $1.06\,\omega_r$ in the shell trap (gray ellipsoid). (b) \emph{In situ} integrated 2D density (in units of $\SI{}{\micro\meter^{-2}}$) of a dynamical ring rotating at Mach $15$, with $2\times10^4$ atoms.
Image taken $\SI{48}{\second}$ after the end of the stirring procedure. The green dashed ellipsoid is a fit of the ring shape, see text for details.
}
\end{figure}

\begin{figure*}[t] 
\includegraphics[width=16cm]{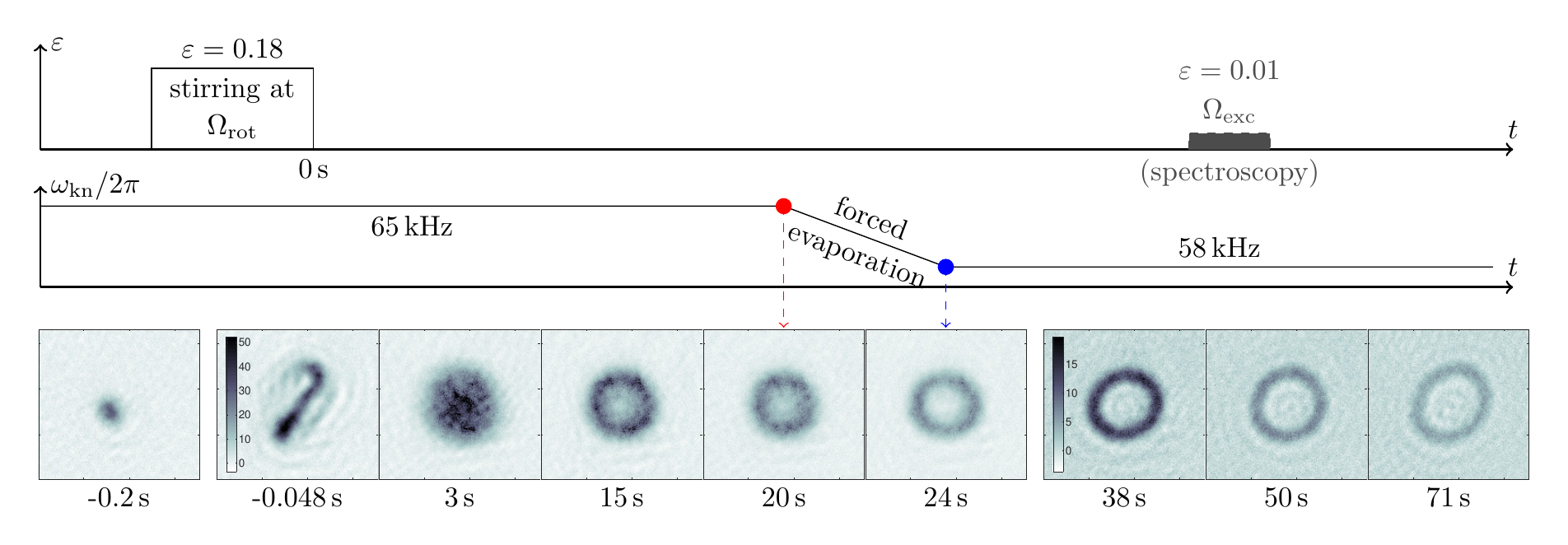}
\caption{\label{fig:movie_time_evolution} 
Sketch of the experimental procedure (see text for detail on the stirring, evaporation, and optional spectroscopy stages) and \emph{in situ} images of the atomic distribution.
The leftmost image shows a cloud at rest before the stirring procedure: only $\sim10\%$ of the atoms are imaged. As the cloud is set into rotation the peak density decreases and we use two different color scales for pictures taken before and after $t=\SI{25}{s}$, for which the darkest pixels correspond to densities of $\SI{50}{\micro\meter^{-2}}$ and $\SI{20}{\micro\meter^{-2}}$, respectively. The three last images correspond to rotations at Mach 11.7, 15.6, and 18.4, respectively.
}
\end{figure*}

In this Letter, we present what is to our knowledge the first experimental realization of such a superfluid annular flow stabilized by its own angular momentum, as shown in Fig.~\ref{ring358}. As the ring shape is directly linked to the atomic motion, we call it ``dynamical ring'' in the following.
We demonstrate that it is a very long-lived quasi-two-dimensional (2D) stable structure that persists over more than a minute. The ring atomic density distribution agrees with a zero-temperature superfluid model. We measure rotation frequencies reaching $1.06\omega_r$ corresponding to a linear supersonic velocity of Mach 18 with respect to the peak speed of sound. We perform the spectroscopy of elementary excitations of the ring and observe the quadrupolelike modes predicted by the diffuse vorticity approach \cite{Cozzini2005}. Above $\Omega_h$, the mode frequencies disagree with this simplified approach.

Exploring the very fast rotation regime up to the formation of a dynamical ring requires a very smooth potential, rotationally invariant around an axis $z$ and anharmonic along the radial coordinate $r=\sqrt{x^2+y^2}$. In previous works we have used radio-frequency (rf) dressed adiabatic potentials~\cite{Garraway2016} to create trapping potentials $V(r,z)$ in the shape of an ellipsoidal surface~\cite{Merloti2013a}, rotationally invariant around the vertical axis $z$. In the presence of gravity, atoms gather at the bottom of this shell. The resulting trap is extremely smooth because of the macroscopic size of the coils creating the potential with respect to the size of the atomic cloud. We have used previously this smoothness and the natural thinness of the shell to bring a condensate in the quasi-2D regime \cite{Merloti2013a} and observe collective excitations~\cite{Dubessy2014,DeRossi2016}.

Here, we take advantage of the smoothness and the weak anharmonicity of this trap to prepare a dynamical ring. Rotational invariance is critical to this aim, and is ensured at the $10^{-3}$ level by a fine-tuning of the dressing field polarization and of the static magnetic field gradients \cite{Perrin2017}.
The trapping frequencies at the bottom of the shell in the harmonic approximation are $\omega_z=2\pi\times\SI{356.5\pm 0.2}{\hertz}$ in the vertical direction and $\omega_r=2\pi\times\SI{33.7\pm 0.04}{\hertz}$ in the horizontal plane, without measurable in-plane anisotropy.
This trap is loaded with a pure BEC of $\num{2.5e5}$ \isotope[87]{Rb} atoms with no discernible thermal fraction.
This atomic cloud has a chemical potential of $\mu/\hbar=2\pi\times\SI{1.8}{\kilo\hertz}$ much greater than $\omega_r$ and $\omega_z$, well in the three-dimensional Thomas-Fermi (TF) regime.
In addition to the dressing field at frequency $\omega_{\rm rf}=2\pi\times\SI{300}{\kilo\hertz}$ and Rabi-coupling $\Omega_{\rm rf}=2\pi\times\SI{48}{\kilo\hertz}$, a radio-frequency knife with frequency $\omega_{\rm kn}$ is used to set the trap depth to approximately $\omega_{\rm kn}-\Omega_{\rm rf}$ by outcoupling the most energetic atoms in the direction transverse to the ellipsoid~\cite{Garrido2006,KollengodeEaswaran2010}.

From this point on, the experiment proceeds as follows: angular momentum is injected into the cloud by rotating the trap. Rotation is further increased by an angular momentum selective evaporation process, resulting in a dynamical ring sustained by its own rotation.
The initial rotation of the cloud is induced by a controlled elliptical deformation of the trap rotating in the horizontal plane at an angular frequency $\Omega_{\rm rot}$. During this stage the potential in the harmonic approximation reads:
\[
V_{\rm rot}(x^\prime,y^\prime)=\frac{M\omega_r^2}{2}\left[(1+\varepsilon)x^{\prime2}+(1-\varepsilon)y^{\prime2}\right]+\frac{M\omega_z^2}{2}z^2,
\]
where $x^\prime=x\cos{(\Omega_{\rm rot}t)}+y\sin{(\Omega_{\rm rot}t)}$ and $y^\prime=-x\sin{(\Omega_{\rm rot}t)}+y\cos{(\Omega_{\rm rot}t)}$ are the coordinates in the rotating frame and $M$ is the atomic mass.
Such deformation of the trap couples to the BEC quadrupole mode and allows us to inject angular momentum into the system.
For a weak ellipticity $\varepsilon$ one expects sharp resonances for vortex nucleation at $\Omega_{\rm rot}=\pm\omega_r/\sqrt{2}$~\cite{Madison2001}. As the ellipticity increases, this resonance is broadened. In this work we consider the extreme case of a large ellipticity $\varepsilon=0.18$, such that the resonance broadening induces a nonzero coupling for $\Omega_{\rm rot}>\sqrt{1-\varepsilon}\,\omega_r$ and therefore rotation is induced by destabilizing the cloud in the weakly trapped direction~\cite{Madison2001,Sinha2005}. Of course the shell trap is not purely harmonic and at this point higher order terms in the confinement potential play a role, such that the atoms stay trapped during the excitation phase \footnote{See the supplemental material, which includes Refs. \cite{Holzmann2008,Holzmann2008a,Prokofiev2001}, for details on potential modelling, the two-dimensional regime and the measurement of the rotation frequency.}.\nocite{Holzmann2008,Holzmann2008a,Prokofiev2001}

The experimental procedure is depicted in Fig.~\ref{fig:movie_time_evolution}. 
We dynamically change the trap geometry by a time-dependent control of the dressing field polarization. Over a time $t_{\rm ramp}=\SI{400}{\micro\second}$ we linearly increase $\varepsilon$ up to its maximal value $0.18$, then rotate the trap axis at angular frequency $\Omega_{\rm rot}=2\pi\times\SI{31}{\hertz}$ for $t_{\rm rot}=\SI{177}{\milli\second}$ and finally restore the isotropic trap over $t_{\rm ramp}$. During this whole process $\omega_r$ and $\omega_z$ are kept constant.
After this procedure we let the cloud evolve in the rotationally invariant trap and take an absorption image of the \emph{in situ} atomic distribution, as reported in Fig.~\ref{fig:movie_time_evolution}.
During the stirring phase the density profile is strongly deformed, as is clear on the second image.
Once the isotropy of the trap is restored, which we take as $t=0$, the cloud shape goes back to circular with an increased radius due to its higher angular momentum. Indeed, in the frame corotating with the atoms at $\Omega$, the effective potential modified by the centrifugal force reads $V_{\rm eff}(r,z)=V(r,z)-M\Omega^2r^2/2$, leading to a reduced effective radial trapping frequency \cite{Note1}.
Because of this size increase, the chemical potential is reduced and the gas enters the quasi-2D regime $\mu\leq\hbar\omega_z$.
After a few seconds a density depletion is established at the center of the cloud which is a signature of $\Omega$ now exceeding $\omega_r$.
Between $t=\SI{20}{\second}$ and $t=\SI{24}{\second}$, we ramp down linearly $\omega_{\rm kn}$ by $2\pi\times\SI{7}{\kilo\hertz}$. After this ramp, a macroscopic hole appears in the profile, indicating that $\Omega$ is now above $\Omega_h$ and that a fast rotating dynamical ring with a typical radius of $\sim\SI{30}{\micro\meter}$ has formed.

To characterize this dynamical ring we measure the effective rotation of the atomic cloud by evaluating the radius at which the peak density occurs and comparing to a model of the full shell trap potential.
The annular density profile presents a small anisotropy, as discussed later in this Letter. For this reason, we fit the peak density along the ring by an ellipse and extract the short and long radii ($r_{\rm short}$ and $r_{\rm long}$ respectively) as well as the orientation.
To evaluate the accuracy of this measurement we compare it to a more direct measurement of the rotation frequency obtained by monitoring the time-of-flight expansion of the dynamical ring, during which the density undergoes a self-similar expansion \cite{Read2003}.
Both methods give similar results, with the same accuracy \cite{Note1}.
The observed anisotropy of the dynamical ring results in a systematic relative uncertainty at the level of $\sim1\%$ in the measurement of $\Omega$.

\begin{figure}[t]
\includegraphics[height=6cm]{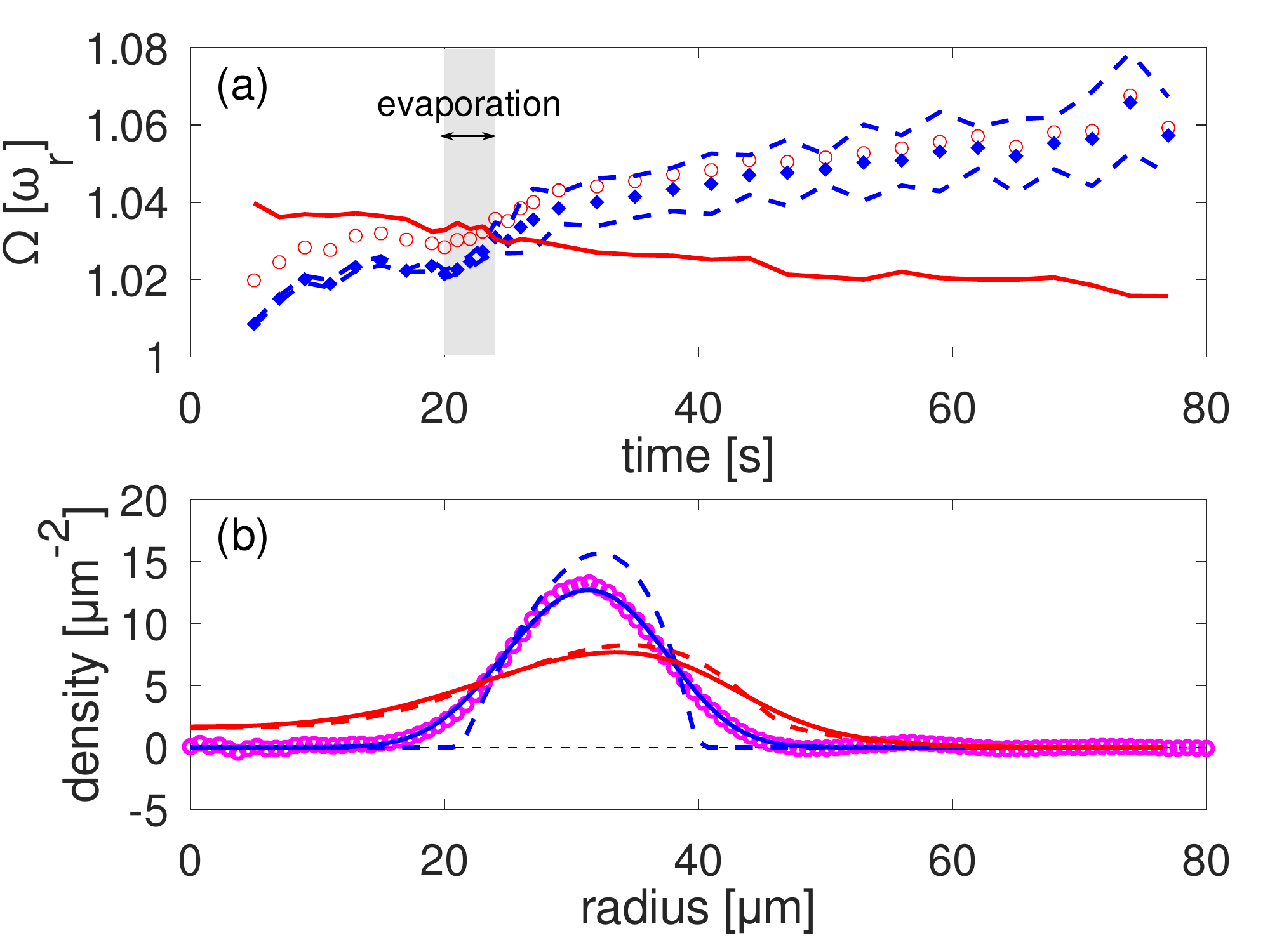}
\caption{\label{fig:Lz} 
(a)
Time evolution of the effective rotation frequency $\Omega$ (symbols) compared to the value of $\Omega_h$ (red solid line).
Filled blue diamonds: $\Omega$ extracted from the fit of the density by an ellipse. Dashed blue lines: associated systematic uncertainty induced by the ring anisotropy. Open red circles: same results taking into account the finite optical resolution, see text for details.
(b) Radial density profile at $t=\SI{35}{s}$ (open magenta circles) compared to two models: a semiclassical Hartree-Fock model at the critical temperature (solid red line) and a TF model (solid blue line). The two models include a 2D convolution with a Gaussian of $\sigma\simeq\SI{4}{\micro\meter}$ to account for the optical resolution.
The dashed lines show the predicted density profiles \emph{before} the convolution.
}
\end{figure}

Figure~\ref{fig:Lz}(a) shows the time evolution of the measured rotation frequency during the experimental sequence.
After the initial stirring phase, the cloud rotation accelerates and reaches a steady state value $\Omega\simeq1.02\omega_r$ around $t=\SI{12}{\second}$ in the presence of a rf knife at $\omega_{\rm kn}=2\pi\times \SI{65}{\kilo\hertz}$.
Applying the forced evaporation phase leads to a significant increase of the measured rotation frequency, as visible in Fig.~\ref{fig:Lz}(a) after $t=\SI{20}{\second}$.
We attribute this to a selective evaporation process of the lower angular momentum states. Indeed, the rf Rabi coupling $\Omega_{\rm rf}$ depends on the position in the shell and is larger for smaller radii \cite{Merloti2013a}, resulting in a smaller trap depth $\omega_{\rm kn}-\Omega_{\rm rf}$ close to the $z$ axis. As low-momentum states have a higher density probability near $z=0$, they are outcoupled more efficiently by the rf knife \cite{Engels2003}. We point out that the 2D collision rate $\Gamma_{\rm coll}$~\cite{Petrov2001} decreases slowly over the time span of the experiment from \SI{30}{\second^{-1}} to \SI{5}{\second^{-1}}, hence being always below $\omega_r$ but large enough to ensure an efficient evaporation.

To confirm the superfluid character of the system the most direct way would be to image the ring after a time-of-flight expansion and look for vortices expected to be present in the bulk of the annulus. This would require a time-of-flight duration long enough for the vortex size to overcome the optical resolution. However, for such a duration the atomic density drops dramatically due to the fast radial expansion and falls below our detection threshold. Instead, we study the \emph{in situ} density profile of the dynamical ring and compare it to two different models of a quasi-2D gas: (i) a semiclassical self-consistent Hartree-Fock model~\cite{Holzmann2008,Holzmann2008a} to approximate the density profile in $V_{\rm eff}(r,z)$ close to the critical temperature and (ii) a zero temperature TF model. 
We point out that it is crucial to convolve the model profiles with a Gaussian of $\sigma\simeq\SI{4}{\micro\meter}$ to account for the imaging resolution and quantitatively reproduce the data~\cite{Note1}.

Figure~\ref{fig:Lz}(b) shows the result of this comparison for a dynamical ring at $t=\SI{35}{s}$ where the two models are adjusted to the experimental profile by fitting the effective rotation rate $\Omega$ and ---for model (i)--- the temperature or ---for model (ii)--- the chemical potential. There are no other free parameters, in particular the trap geometry is fixed by an independent measurement. We find that, for all the pictures which present a density depletion at the center of the cloud (i.e., for $t\geq\SI{7}{s}$), the convoluted TF profile is better at reproducing the experimental density profile than the semiclassical profile. In particular model (i) does not capture the full density depletion at the center at $t\geq\SI{25}{s}$ and does not reproduce the measured peak density. On the contrary these two features are correctly captured by the TF model. We therefore conclude that our samples are well below the degeneracy temperature. This analysis shows that the finite imaging resolution leads to a small systematic underestimation of the rotation frequency by $\sim1\%$ when it is measured using the ellipsoid radii.

Using the Thomas-Fermi model we estimate the properties of the cloud. For example the TF profile shown on Fig.~\ref{fig:Lz}(b) has a chemical potential of $\mu/\hbar\simeq2\pi\times\SI{84}{\hertz}$ and an averaged angular momentum per particle $\braket{L_z}/N\simeq\hbar\times317$. Interestingly the estimated peak speed of sound $c=\sqrt{\mu/M}\simeq\SI{0.62}{mm/s}$ at the peak radius $r_{\rm peak}$ is much smaller than the local fluid velocity $v=\Omega r_{\rm peak}\simeq\SI{6.9}{mm/s}$: the superfluid is therefore rotating at a supersonic velocity corresponding to a Mach number of 11 \footnote{Very recently supersonic velocities were also achieved for trapped atoms accelerated along an annular guide \cite{Pandey2019}.}.\nocite{Pandey2019}
Moreover, due to the continuous acceleration of the rotation, the dynamical ring radius grows gradually with time, which results in a decrease of the chemical potential and an increase of the Mach number.
For $t>\SI{45}{s}$ the chemical potential is below $2\hbar\omega_r$ and the highest measured Mach number is above 18.

One can observe in the density profiles of Fig.~\ref{fig:movie_time_evolution} that the ring is not circular at all times. This anisotropy develops after the forced evaporation stage and rotates once established, in a way reminiscent of a quadrupole surface mode for a connected hydrodynamic gas~\cite{Zambelli1998}. For a connected cloud rotating at $\Omega<\Omega_h$, the energies of the $m=\pm 2$ quadrupole modes are nondegenerate and their difference allows us to measure $\Omega$~\cite{Chevy2000,Haljan2001b}. Such quadrupole modes have also been predicted for a dynamical ring formed in a harmonic-plus-quartic trap~\cite{Cozzini2005} but have not been observed up to now.
We have investigated these modes in our system by using a surface mode spectroscopy scheme~\cite{Bretin2003}.
We  selectively excite a quadrupole mode by rotating a small trap anisotropy  $\varepsilon=0.01$ for a duration $\tau=\SI{1}{\second}$ once the ring is formed. For each excitation frequency $\Omega_{\rm exc}$, the cloud is imaged \emph{in situ} just after excitation and the cloud anisotropy $\zeta = r_{\rm long}/r_{\rm short}$ is plotted as a function of $\Omega_{\rm exc}$.

\begin{figure}[t]
\includegraphics[width=\linewidth]{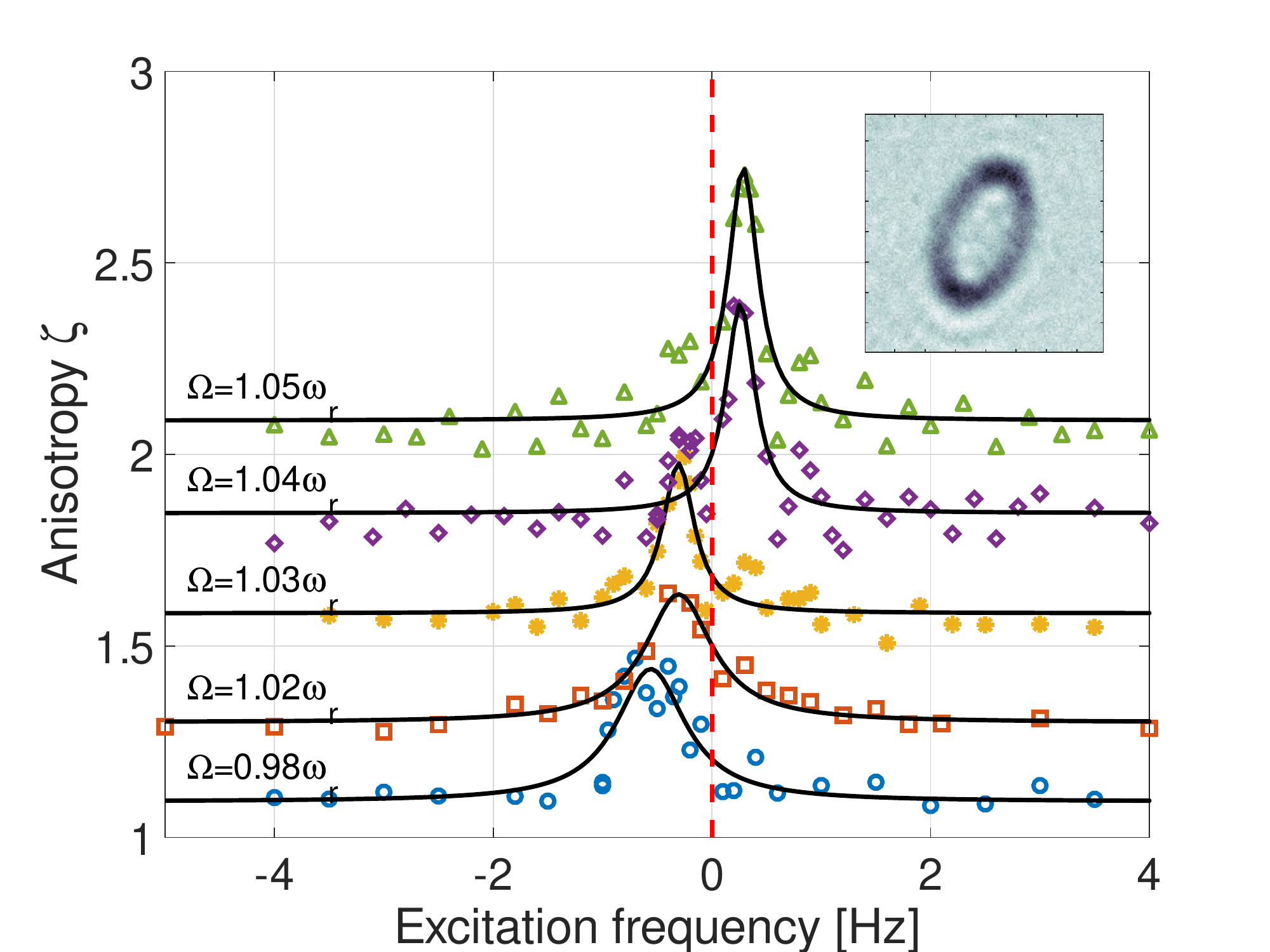}
\caption{\label{fig:anis_frot}
Cloud anisotropy $\zeta$ as a function of the probe frequency $\Omega_{\rm exc}$ for different times 
$t=$\SIlist{2.5;5;20;26;50}{\second} from bottom to top,
corresponding to effective rotation rates $\Omega/\omega_r\simeq 0.98,1.02,1.03,1.04$, and $1.05$ (circle, square, star, diamond, and triangle symbols, respectively). The black solid curves are Lorentzian fits to the data.
Inset: example of resonantly excited dynamical ring.
}
\end{figure}

Figure~\ref{fig:anis_frot} shows the result of this quadrupole mode spectroscopy, focusing on the $m=-2$ mode, for increasing rotation frequencies. 
By convention we set that $\Omega_{\rm exc}$ is negative when the excitation anisotropy is rotated against the direction of the flow. We find that the $m=-2$ quadrupole mode resonance occurs at negative values of $\Omega_{\rm exc}$ for $\Omega<\Omega_h$, as expected. However, when the rotation frequency increases and $\Omega>\Omega_h$ this resonance shifts towards positive frequencies, meaning that this mode is now corotating with the flow. This is not expected within the diffuse vorticity approach~\cite{Cozzini2005} which in our case always predicts a negative value for the $m=-2$ resonant quadrupole frequency. This  provides an explanation to the observed ellipticity of fast-rotating dynamical rings: during the forced evaporation phase where the flow accelerates, the mode frequency vanishes and at this point any residual \emph{static} anisotropy of the trap excites resonantly this mode. 
We observe that this ring anisotropy is corotating with the atomic flow at a very low angular velocity and remains observable for \SI{10}{\second}.

In conclusion, we have demonstrated the first experimental realization of a superfluid dynamical ring, an important step towards the observation of a giant vortex. This state should be accessible in our experimental system for an atom number of 400 atoms \cite{Kasamatsu2002}, below our current detection sensitivity but well within reach of single atom detection schemes \cite{Buecker2009}.

Owing to the smoothness of the rf-dressed adiabatic trap, the rotation is preserved over a minute, even if the atomic velocity is more than ten times the superfluid speed of sound. One could wonder how a localized defect would dissipate a superfluid flow at such a supersonic speed. This would complete to even higher speeds the theoretical and recent experimental works~\cite{Law2000,Pavloff2002,Paris-Mandoki2017,Bradley2016,Dries2010} that have shown that obstacles moving at velocities far exceeding the Landau critical velocity do not necessarily create a significant amount of excitations.

Finally, we have revealed the existence of weakly damped collective quadrupole modes of the dynamical ring. At very fast rotation, the observed frequency of the low frequency mode does not agree with hydrodynamic, diffuse vorticity calculations \cite{Cozzini2005}.
This suggests the need of more refined theoretical models beyond the diffuse vorticity approximation and stimulates further experimental investigation of the excitation spectrum.
The creation of this rotating state offers fascinating perspectives for the study of supercritical superfluid flows~\cite{Kasamatsu2002,Bradley2016,Paris-Mandoki2017}.

\acknowledgments
We thank J. Walraven, A. Minguzzi, F. Chevy, S. Stringari and J. Beugnon for helpful discussions and F. Wiotte and A. Arnold for their contribution to the rf control of the shell trap. We acknowledge financial support from the ANR project SuperRing (Grant No. ANR-15-CE30-0012) and from the R\'egion \^Ile-de-France in the framework of DIM SIRTEQ (Science et Ing\'enierie en R\'egion \^Ile-de-France pour les Technologies Quantiques), project DyABoG.

%

\clearpage
\begin{center}
{\large \textbf{Supplemental material}}
\end{center}

\setcounter{section}{0}
\section{Adiabatic potential\\in the rotating frame}
\paragraph{Shell trap}
Rubidium 87 atoms are confined in their ground state $F=1$ of Land\'e $g$-factor $g_F$ in the shell-like adiabatic potential fully discussed in Ref. \cite{Merloti2013a}. The atoms are placed in a quadrupole magnetic field of symmetry axis $z$ and in-plane gradient $\hbar\alpha/|g_F|$ and dressed by a radio-frequency (rf) field of maximum coupling $\Omega_{\rm rf}$ at the bottom of the shell. In a frame rotating at frequency $\Omega$, the effective dressed trap potential taking into account the centrifugal potential reads:
\[
V_{\rm eff}(r,z)=\hbar\sqrt{(\alpha\ell-\omega)^2+\Omega_c(r,z)^2}+Mgz-\frac{1}{2}M\Omega^2r^2
\]
where $\Omega_c(r,z)=\frac{\Omega_{\rm rf}}{2}\left(1+\sqrt{1-(r/\ell)^2}\right)$ is the local rf Rabi coupling, $\ell=\sqrt{r^2+4z^2}$, and $(r,z)$ are the cylindrical coordinates. $\hbar$ is the reduced Planck constant, $g=9.81$~m/s$^2$ is the normal gravitational acceleration and $M$ is the atomic mass.
The energy difference between dressed states is $\hbar\Omega_c(r,z)$ and decreases with $r$.

The equilibrium properties in the absence of rotation ($\Omega=0$) are well known~\cite{Merloti2013a}: the minimum of the trapping potential is located at $r=0$ and $z=z_0$ with
\[
z_0=-\frac{\omega}{2\alpha}\left(1+\frac{\epsilon}{\sqrt{1-\epsilon^2}}\frac{\Omega_{\rm rf}}{\omega}\right),
\]
where $\epsilon=Mg/(2\hbar\alpha)$ is a small parameter accounting for the gravitational sag.
Around this equilibrium position the potential is locally harmonic with frequencies:
\[
\omega_z=2\alpha\sqrt{\frac{\hbar}{M\Omega_{\rm rf}}}(1-\epsilon^2)^{3/4}
\]
and
\[
\omega_r=\sqrt{\frac{g}{4|z_0|}\left(1-\frac{\hbar\Omega_{\rm rf}}{2Mg|z_0|}\sqrt{1-\epsilon^2}\right)}.
\]
This harmonic approximation is valid for $|z-z_0|,r\ll\omega/\alpha$. Up to fourth order, the radial confinement reads $\frac{1}{2}M\omega_r^2r^2\left(1+\lambda r^2/a_r^2\right)$ where $a_r=\sqrt{\hbar/(M\omega_r)}$ is the size of the harmonic ground state and $\lambda=1.5\times 10^{-4}$ describes the weak anharmonicity.

\begin{figure}[ht]
\includegraphics[width=8cm]{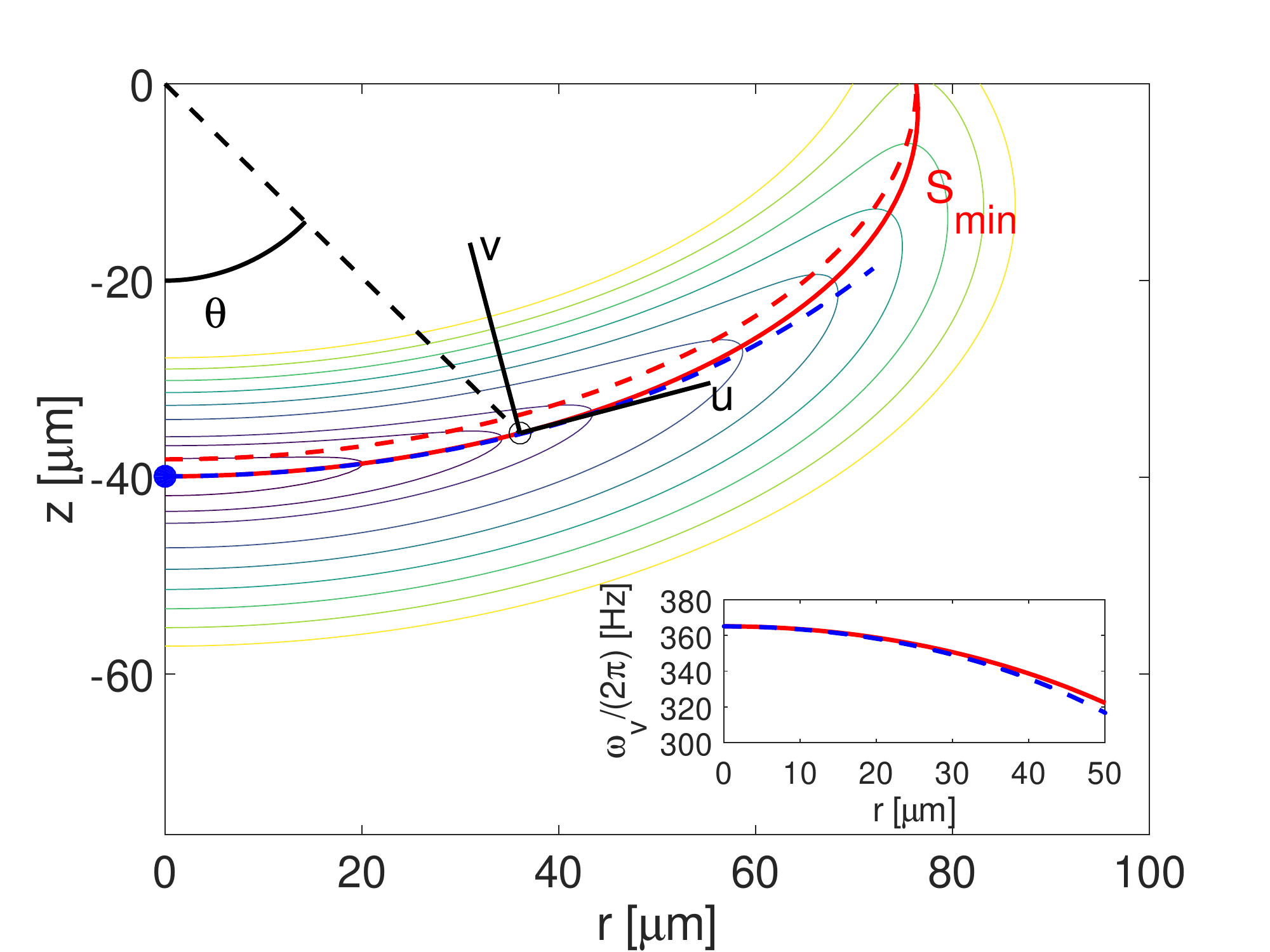}
\caption{\label{FigApp1}Dressed trap iso-potentials in the absence of rotation (with values $\{2,4,6,8,10,20,30,40,50,60,70\}$ kHz above the trap minimum). The equilibrium position is at the bottom (blue disk). Dashed red curve: resonant radio-frequency ellipsoid $\ell=\omega/\alpha$. Solid red curve: minimal force surface $S_{\rm min}$ for each angle $\tan{\theta}=-r/z$. Dashed blue line: locus of the trap minimum position in a rotating frame when $\Omega/(2\pi)$ spans $[0,50]$ Hz.
Inset: Solid red curve: effective transverse confinement frequency of the \textit{static} trap along the minimal force surface in the direction normal to this surface; Dashed blue curve: same transverse frequency for a \textit{rotating} trap at the trap equilibrium position.
}
\end{figure}

For $\Omega<\omega_r$ the equilibrium position remains on the axis $r=0$ at $z=z_0$, and the only difference is a renormalization of the radial trapping frequency: $\omega_r^{\rm eff}=\sqrt{\omega_r^2-\Omega^2}$. Of course, as the frequency decreases, the trap anharmonicity becomes more important in the determination of the cloud shape.

For $\Omega>\omega_r$ the trap minimum is located at a non zero radius, just below the surface $S_{\rm min}$ defined by $\ell=\omega/\alpha$ (corresponding to the resonant radio-frequency), due to the combined gravitational and rotational sags. There is no simple analytic expression available in this case.

In the experiment we observe that for $\Omega\sim\omega_r$ the cloud extends beyond the region where the harmonic approximation is valid.
It is therefore necessary to study how the trap properties evolve with the rotation frequency. We perform a numerical computation with our experimental parameters: $\omega=2\pi\times 300$~kHz, $\Omega_{\rm rf}=2\pi\times48.17$ kHz, $\alpha=2\pi\times3.93$ kHz/$\mu$m and $M=1.44\times10^{-25}$ kg.

\paragraph{Effective coordinates}
The trap minimum is displaced for $\Omega>\omega_r$ but remains very close to the ``minimal force surface'' $S_{\rm min}$, as seen in Fig.~\ref{FigApp1}, for a given vertical plane. This surface is defined by finding for each angle $\theta$ the distance $\rho_0(\theta)$ at which the norm of the force derived from $V_{\rm eff}(r,z)$ is minimal, along the ray defined by the angle $\theta$. For each angle $\theta$ one can introduce the angle $\phi(\theta)$ between the tangent plane to $S_{\rm min}$ and the horizontal plane and define the local curvature of the potential in the direction perpendicular to $S_{\rm min}$ associated to an effective transverse confinement $\omega_v(\theta)$. Inside $S_{\rm min}$ one can define an effective two-dimensional, rotationally invariant, trapping potential $V_{\rm surf}(\theta)=V(\rho_0(\theta)\sin{\theta},-\rho_0(\theta)\cos{\theta})$.
The inset in Fig~\ref{FigApp1} show that the effective rotating trap potential is well approximated by:
\begin{equation}
\label{eqn:approx}
V_{\rm eff}=V_{\rm surf}(\theta)-\frac{M}{2}\Omega^2\sin{[\phi(\theta)]}^2\rho_0(\theta)^2+\frac{M}{2}\omega_v(\theta)^2v^2,
\end{equation}
where $v$ is the coordinate along the direction normal to the minimal force surface, at the angle $\theta$.

We can now introduce an effective potential for $v$ and the curvilinear coordinate $u$ instead of the $(r,z)$ coordinates, using a mapping $\theta\to u$. The potential reads:
\begin{equation}
\label{eqn:Vq2D}
\tilde{V}_{\rm eff}(u,v)=V_{\Omega}(u)+\frac{M}{2}\omega_v(u)^2v^2,
\end{equation}
where $V_{\Omega}(u)=V_{\rm surf}(u)-\frac{M}{2}\Omega^2\sin{[\phi(u)]}^2\rho_0(u)^2$. As the potential $V_{\Omega}(u)$ varies smoothly and $\hbar\omega_v$ is larger than the chemical potential already right after the stirring phase, this justifies a quasi two-dimensional treatment in the rotationally invariant potential $V_\Omega(u)$. $V_\Omega(u)$ is minimal for $u=u_{\rm eq}$ which is non zero for $\Omega>\omega_r$.

\section{Quasi two-dimensional model}
To compute thermodynamic quantities for the atomic gas, we rely on the semi-classical quasi-two-dimensional model developed in Refs.~\cite{Holzmann2008,Holzmann2008a}, using the local density approximation (LDA) to include the effective 2D trap potential in the rotating frame. Here we take into account the rotation in the effective trap potential but neglect its effect as a gauge field on the kinetic energy operator.
We summarize here the main results of Refs.~\cite{Holzmann2008,Holzmann2008a} that we use.
Given a separable potential, as in Eq.~\eqref{eqn:Vq2D}, one can approximate the quasi two-dimensional phase space density $D(u)=n(u)\lambda_T^2$ by:
\begin{equation}
\label{eqn:D2D}
D(u)=-\sum_{\nu=0}^\infty\ln{\left[1-e^{\beta(\mu_{\rm loc}(u)-\nu\hbar\omega_v)}\right]},
\end{equation}
where $n(u)$ is the 2D density $\beta=1/(k_BT)$ is the inverse temperature, $\lambda_T=\sqrt{2\pi\hbar^2\beta/M}$ is the thermal de Broglie wavelength and, within LDA, we define:
\[
\mu_{\rm loc}(u)=\mu-2g_2n(u)-V_{\Omega}(u)+\left[V_{\Omega}(u)\right]_{u=u_{\rm eq}},
\]
so that the chemical potential $\mu$ is defined for a trap with vanishing potential at the equilibrium position. 
$g_2$ is the effective two-dimensional interaction constant:
\[
g_2=\frac{4\pi\hbar^2a_{\rm sc}}{M}\int dv[\rho(v,v)]^2=\sqrt{8\pi}\frac{a_{\rm sc}}{a_v}\frac{\hbar^2}{M}\sqrt{\tanh{\left[\frac{\hbar\omega_v}{k_BT}\right]}},
\]
where $a_{\rm sc}$ is the three-dimensional s-wave scattering length and $\rho(v,v)$ is the diagonal of the normalized density matrix in the transverse direction and the last equality holds for a transverse harmonic confinement of typical length $a_v=\sqrt{\hbar/(M\omega_v)}$. Here the effective two-dimensional interaction strength is re-normalized only by the temperature and not by the mean-field interactions (which is accurate for $\mu \leq \hbar\omega_v$).

\begin{figure}[t]
\includegraphics[width=8cm]{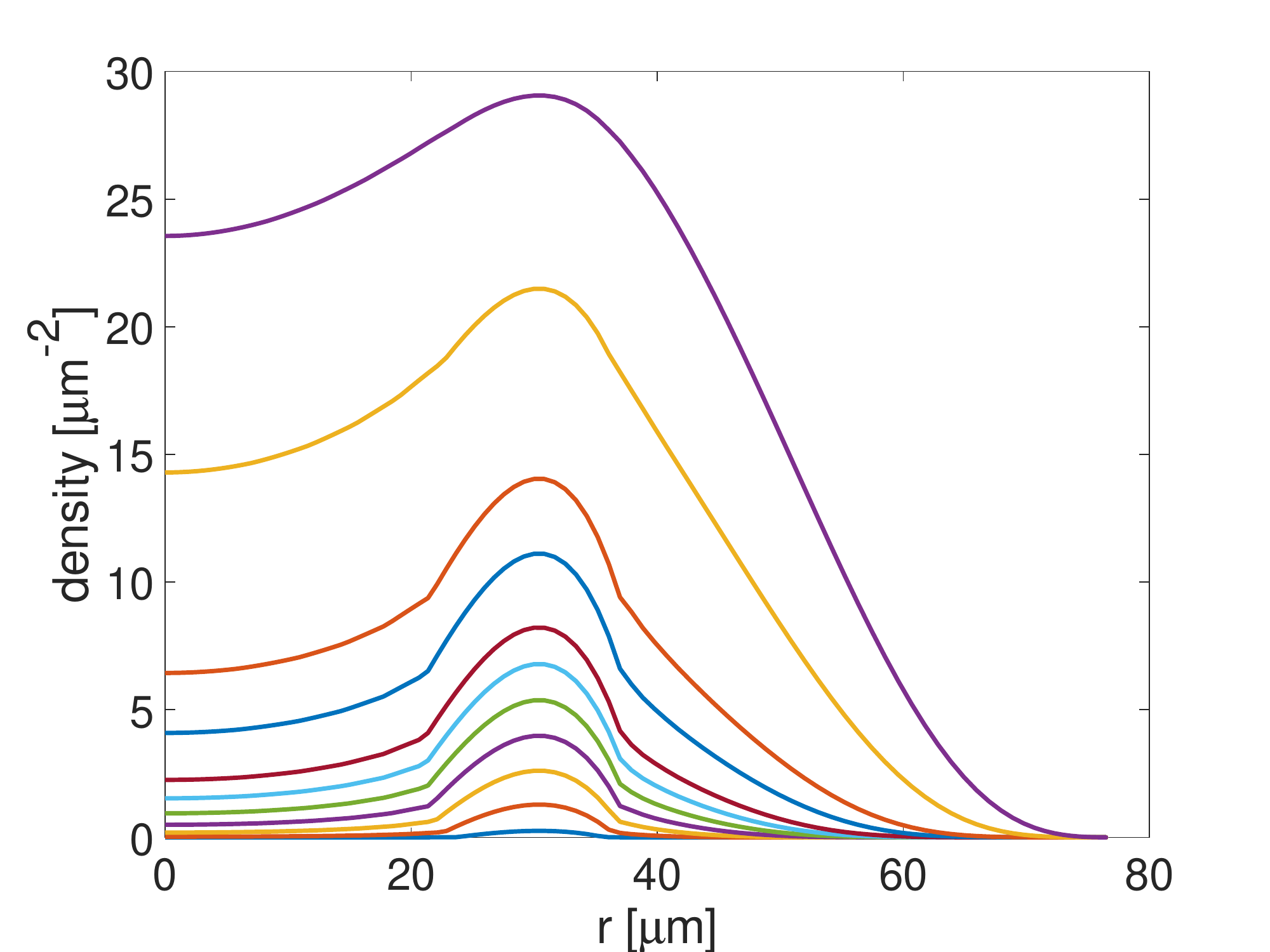}
\caption{\label{FigApp2}Equilibrium density radial profiles at the critical temperature $T$ spanning the range $\{1,5,10,15,20,25,30,40,50,75,100\}$~nK (from bottom to top) and a rotation of $\Omega/(2\pi)=35.5$~Hz. }
\end{figure}

Equation~\eqref{eqn:D2D} describes the phase space density accurately below the critical point for the Berezinskii--Kosterlitz--Thouless (BKT) transition \cite{Prokofiev2001}:
\[
D\leq D_c=\ln{\left[\frac{380\hbar^2}{Mg_2}\right]},
\]
and above one can use an approximate Thomas-Fermi model:
\[
n(u)=\frac{\mu-V_{\Omega}(u)}{g_2},~\textrm{for}~n(u)>D_c\lambda_T^{-2}.
\]
In our case $g_2$ weakly depends on $u$ because $\omega_v$ itself depends on $u$, see Fig.~\ref{FigApp1}. For the sake of simplicity we will assume that it is constant and equal to the non rotating value.

We can now compute the critical atom number for the BKT superfluid transition as a function of the temperature, for example in a frame rotating at $\Omega/(2\pi)=\SI{35.5}{\hertz}\simeq1.05\omega_r$, as well as density profiles at the transition, as shown in Fig.~\ref{FigApp2}.
We can infer from such profiles, in particular by looking at the residual density at the center $u=0$, that in the experiment the atomic cloud must be highly degenerate. For example in Fig.~3(b) of the main paper, the central density is well below $\SI{1}{\micro\meter^{-2}}$ while the peak density is of the order of $\SI{15}{\micro\meter^{-2}}$.

\begin{figure}[t]
\includegraphics[width=8cm]{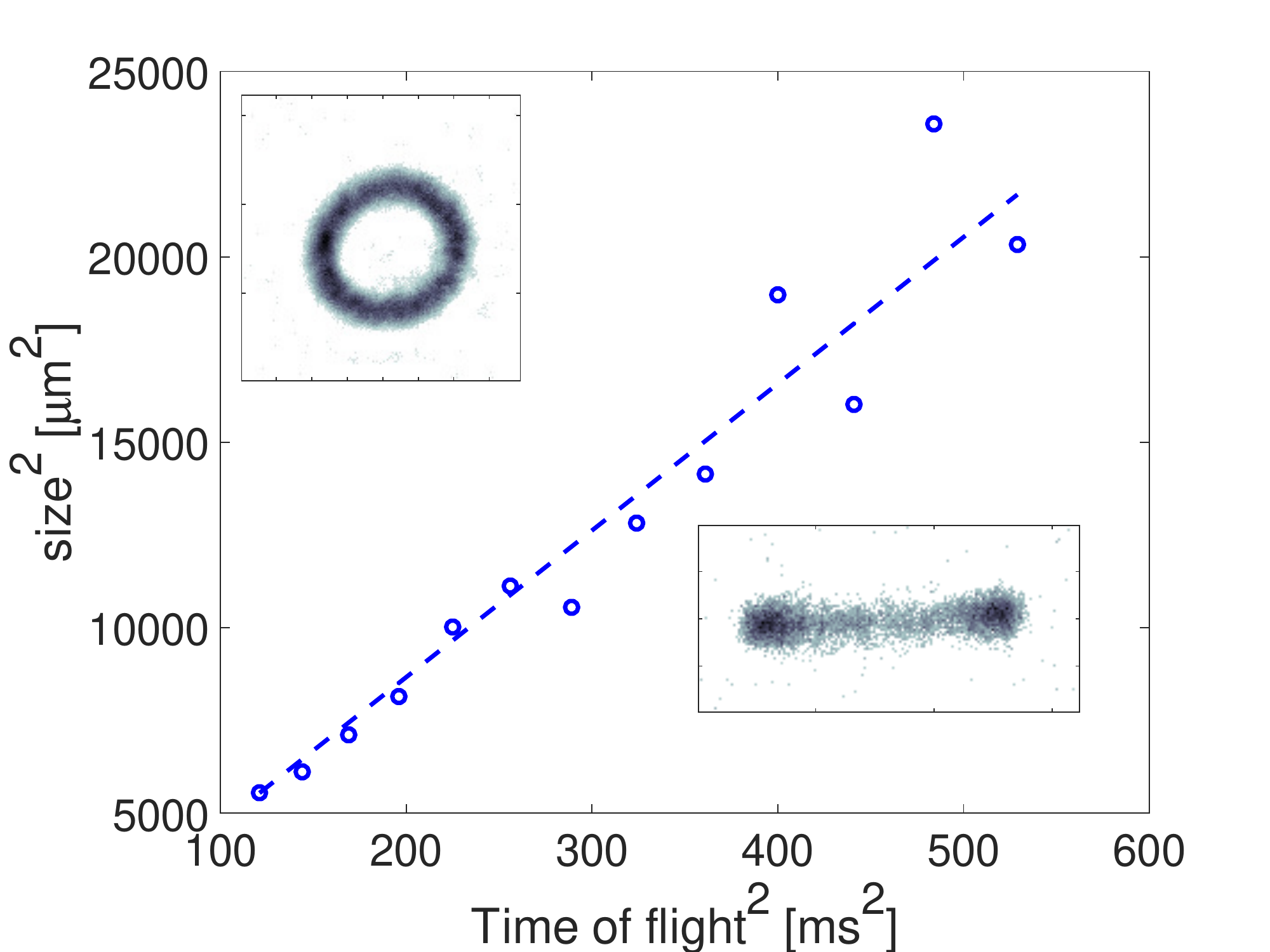}
\caption{\label{FigApp4}
Time-of-flight expansion dynamics: measured $<x^2>$ as a function of the time-of-flight $t_{\rm tof}^2$ (open blue circles). The dashed line results from a single parameter fit of the form $<x_0^2>+(\Omega\times t_{\rm tof})^2$, where $<x_0^2>$ is the \emph{in situ} size independently measured (see top left inset, top view). The bottom right inset shows a typical time-of-flight image (side view) at $t_{\rm tof}=\SI{23}{ms}$.
}
\end{figure}

\section{Measurement of $\Omega$}
We have several means to evaluate the effective angular velocity of the atomic flow. The more direct is to perform a time-of-flight experiment, see Fig.~\ref{FigApp4}: we abruptly turn off the confinement potential and let the atoms expand freely for a time $t_{\rm tof}$. The cloud undergoes a fast radial expansion from which $\Omega$ can be inferred using a ballistic model. The vertical expansion is very fast such that we do not expect a significant contribution of the interactions to the time-of-flight expansion.

\begin{figure}[h]
\includegraphics[width=6.5cm]{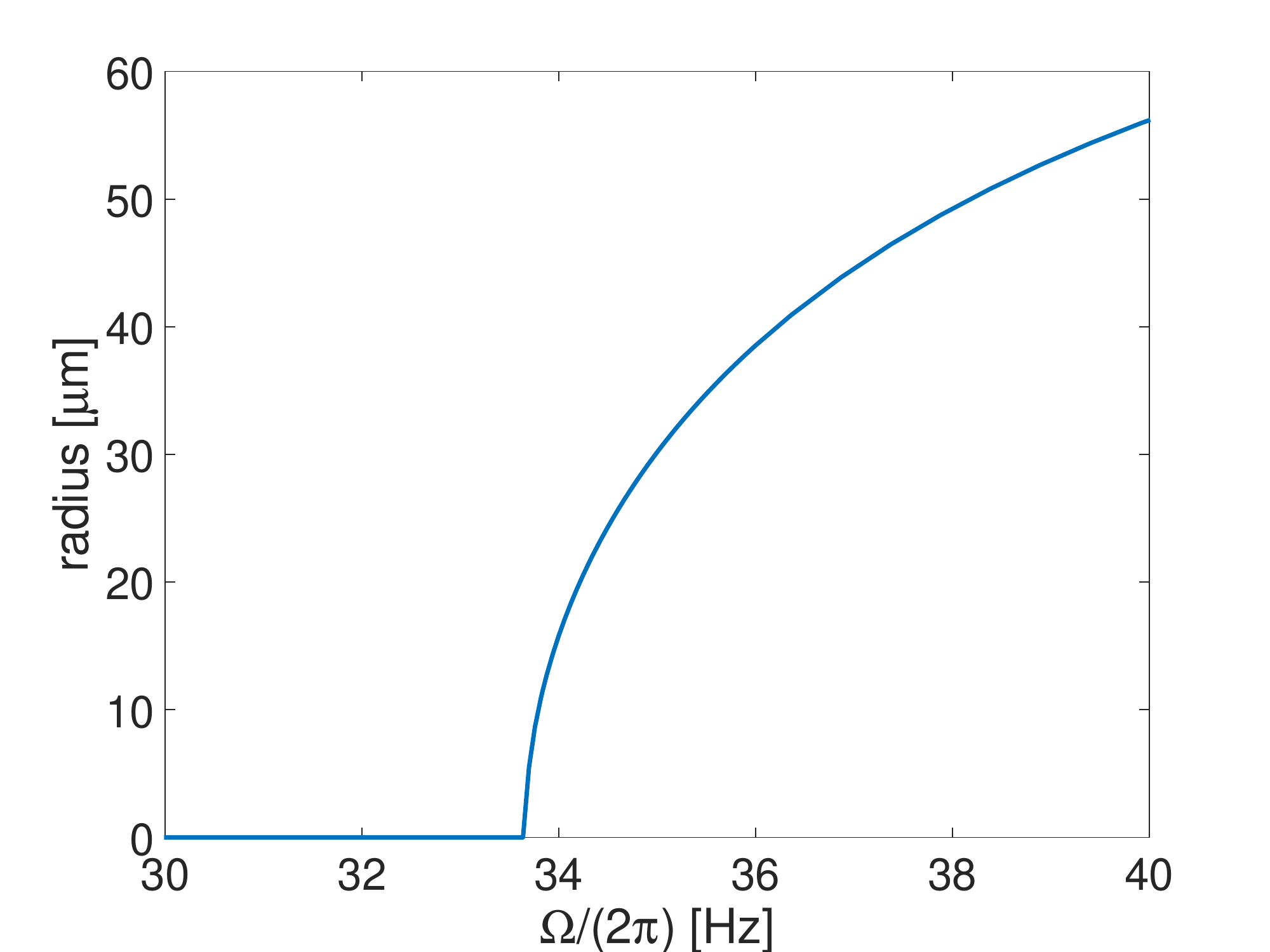}
\caption{\label{FigApp3}Radius of the dynamical ring as a function of the rotation frequency, computed numerically from the the full shell potential.
}
\end{figure}

We also extract $\Omega$ directly from the \emph{in situ} density profiles.
To this aim, we first fit the annular density profile with an ellipse, and from its parameters deduce the position of the trap center, the mean radius, the anisotropy of the dynamical ring and its orientation with respect to the camera axes. 
We then use our knowledge of the adiabatic potential in the rotating frame to derive the effective rotation frequency from the mean radius, assuming that the maximum atomic density occurs at the effective trap minimum, using the relation sketched in Fig.~\ref{FigApp3}.
The values obtained by the two methods are in excellent agreement.

Finally, we also compare directly the experimental density profiles to the quasi-2D model discussed above. This comparison requires to steps: we first perform an azimuthal integration of the density profiles to obtain the mean radial profile of the dynamical ring and we then fit the models to this profile, including a convolution with a Gaussian point spread function of $\sigma=\SI{4}{\micro\meter}$ to take into 
account the image resolution. Before computing the azimuthal integration we correct the small apparent ellipticity of the dynamical rings by appropriately rescaling the picture at fixed average radius. As discussed in the paper this analysis shows that the blurring of the density profiles due to the optical resolution induces a systematic underestimation of the rotation frequency by about 1\%.

\end{document}